\documentstyle[12pt]{article}
\newcommand{\no}{\nonumber\\}
\newcommand{\be}{\begin{equation}}
\newcommand{\ee}{\end{equation}}
\newcommand{\ba}{\begin{eqnarray}}
\newcommand{\ea}{\end{eqnarray}}

\newcommand{\la}[1]{\label{#1}}

\def\gl#1{(\ref{#1})}
\def\tr#1{\mbox{\rm tr}\left(#1\right)}

\renewcommand{\thefootnote}{\fnsymbol{footnote}}
\date{}

\newcommand {\gll}{\gamma_L}
\newcommand {\gs}{\gamma_S}

\newcommand {\dm}{\Delta m}
\newcommand {\dl}{\Delta \lambda}

\newcommand {\Rep}{{\rm Re}\epsilon}

\newcommand {\Rdk}{{\rm Re}\delta_K}
\newcommand {\Idk}{{\rm Im}\delta_K}

\newcommand {\xtt}{x_{33}}

\newcommand {\El}{e^{-\bar{\gamma}_L t}}
\newcommand {\Es}{e^{-\bar{\gamma}_S t}}
\newcommand {\E}{e^{-\frac{1}{2} \gamma t}}

\begin{document}

\begin{center}

{\Large\bf Neutral Kaons in Medium: Decoherence Effects}

\vspace{4mm}
 A. A. Andrianov\footnote{Permanent address: 
Department of Theoretical Physics, St.Petersburg State University,
 198904 St.Petersburg,  Russia}, 
 J. Taron and
 R.Tarrach \\
  Departament d'ECM,
Universitat de Barcelona\\
 08028 Barcelona, Spain\\
and\\
IFAE
\end{center}

\vspace{1cm}

\begin{abstract}
We consider departures from hamiltonian dynamics in the evolution of neutral
kaons due to their interactions with environment that generate entanglement
among them. We propose a phenomenological model of stochastic re-scattering
and estimate the coefficients of the effective hamiltonian and decoherence
terms. Finally we analyze the interplay between the weak interaction and matter
effects and propose the observables suitable to measure the matter
characteristics of CPT violation and decoherence.

\end{abstract}
\bigskip
\renewcommand{\thefootnote}{\arabic{footnote}}
\section{Introduction}
Since the fifties the system of neutral kaons has provided us with a unique
laboratory to investigate the superposition principle, interference
effects and CP violation \cite{BS}, \cite{maiani}.
CP violation was first measured in \cite{CCFT} long ago, 
in the mixing of the neutral 
kaons, the so called {\it indirect} CP violation, 
whereas it was not until recently that {\it direct}
violation, in the decay amplitudes to two pions, was
experimentally established (see \cite{CPLEAR}, \cite{KTEV}, \cite{NA48}).
There are at present a few dedicated experiments
in project to make precision measurements of the CP breaking parameters
\cite{KTEV}, \cite{NA48}, \cite{KLOE}.
The analysis of their data may require to take properly into account the
effects of {\it decoherence} due to entanglement of the neutral kaon
system with the environment in which it evolves, which entails a pure
kaon state to convert into a mixed one. 
We consider the evolution
of neutral kaons in the presence of matter, in an environment that is not the
perfect vacuum, which generates departures from hamiltonian dynamics
in the kaon sub-system. These are effects that exist in addition to the
weak interactions and are dominated by the strong interactions of the kaons
with the environment: they lead to an effective breaking of charge
conjugation, because the environment is made of matter, not of anti-matter;
and to an effective breaking of time-reversal invariance associated to the
large number of degrees of freedom in the environment that leads to 
irreversibility.

Similar analysis can be found in the literature 
\cite{EHNS}, \cite{EMN}, \cite{PH}, \cite{BF}. There, the 
motivation is to study decoherence effects that come from quantum gravity.
Ours is the interaction with the environment that is not the vacuum
\cite{E}.

In section 2 we explain how the propagation in an environment being
projected to the kaon system is described by a Lindblad master equation
\cite{L}
which characterizes the departure from hamiltonian evolution; we introduce
a criterion to quantify decoherence and build a phenomenological
model based on a stochastic re-scattering of kaons in a dense matter
induced by strong interactions. In section 3 we restrict the general analysis
to a two-component $K^0-\bar{K}^0$ system and estimate coefficients of
both effective hamiltonian and Lindblad interaction. In particular we 
conjecture that the decoherent part may be equally important for both 
hamiltonian and Lindblad piece. In section 4
we analyze the interplay between the weak interaction and matter effects
and propose the observables suitable to disentangle the parameter
responsible for CPT violation in matter $\delta_K$, the characteristic of
indirect CP violation in vacuum $\epsilon$, and the decoherence parameter
$x_{33}$. In the appendix we propose a toy model of two level environment
to explain how the Lindblad interaction comes about, in the Markovian 
approximation.

The neutral kaon evolution in the vacuum is analyzed with the Schr\"odinger
equation
\begin{equation}
i \dot{\rho}(t)= H_w \rho(t) - \rho(t) H_w^\dagger,
\label{schrodinger}
\end{equation}
where $H_w$ is 
the LOY \cite{loy} 
effective hamiltonian $H_w=\sum_{I=S,L} \lambda_I 
|K\rangle \langle \tilde{K}_I|$, which is not hermitean and its eigenstates
$\lambda_I= m_I - \frac{i}{2} \gamma_I$ possess a non-vanishing imaginary
part of their decay rates. $S,L$ stand for the {\it short} and {\it long}
components, and $|\tilde{K}_I \rangle$ are such that 
$\langle \tilde{K}_I | K_J \rangle = \delta_{IJ}$. 
Recall \cite{PDG} that $m_S \simeq m_L \simeq 500 \; MeV$, 
$\Delta m = m_L-m_S =3.5 \times 10^{-12} \; MeV$, 
$\gamma_S=7.3\times 10^{-12} \; MeV$ and 
$\gamma_L=1.3 \times 10^{-14} \; MeV$. 

The $|K_I\rangle$ have
diagonal evolution $|K_I(t)\rangle= e^{-i \lambda_I t}|K_I(0)\rangle $. In
terms of the CP eigenstates
\begin{equation}
|K_S\rangle=\frac{1}{\sqrt{1+|\epsilon_S|^2}}\left( |K_1\rangle +
\epsilon_S |K_2 \rangle \right), \;\;\;\;\;
|K_L\rangle=\frac{1}{\sqrt{1+|\epsilon_L|^2}}\left( |K_2\rangle +
\epsilon_L |K_1 \rangle \right);
\label{def}
\end{equation}
they reduce to the CP eigenstates
$|K_1 \rangle=\frac{1}{2} \left( |K^0\rangle + |\bar{K}^0 \rangle \right)$, 
$|K_2 \rangle=\frac{1}{2} \left( |K^0\rangle - |\bar{K}^0 \rangle \right)$,
if CP is conserved.

At this point we recall that CPT conservation requires 
$\epsilon_S=\epsilon_L$; T invariance, $\epsilon_S+\epsilon_L=0$;
and CP, $\epsilon_S=\epsilon_L=0$.

Indirect CP violation has been measured \cite{CCFT}, \cite{PDG}
\begin{equation}
\epsilon \simeq 2.3 \times 10^{-3} e^{i \pi/4},
\label{epsilon}
\end{equation}
(CPT conservation is assumed, $\epsilon_S=\epsilon_L\equiv \epsilon$.)
Direct CP violation has also been measured \cite{KTEV}, \cite{NA48}
\begin{equation}
\epsilon' /\epsilon \sim 2 \times 10^{-3},
\label{epsilonprima}
\end{equation}
where
$$\frac{A(K_L \to \pi^+ \pi^-)}{A(K_S \to \pi^+ \pi^-)} \equiv
\epsilon_L + \epsilon', \;\;\;\;
\frac{A(K_L \to \pi^0 \pi^0)}{A(K_S \to \pi^0 \pi^0)} \equiv
\epsilon_L - 2 \epsilon'\;\;\;, \epsilon_L=\epsilon.$$

\section{Propagation in environment}
The propagation in an environment is dominated by the strong interactions,
and we postpone to the next section the inclusion of weak interactions.
The evolution of the {\it total} system [Kaon + (Large) Environment]
is unitary, with the time evolution operator 
given by $U(t)=\exp(-i H_{Total}t)$;
whereas the dynamics of the kaon sub-system alone is obtained by
tracing out the environment degrees of freedom:
\begin{equation}
\rho_K(t)= {\rm Tr}_{ENV} \left( U(t) \rho_{K+ENV}(0) U^\dagger (t) \right).
\label{evolution}
\end{equation}
This gives a complicated evolution for $\rho_K$, even if the initial conditions
are a direct product $\rho_{K+ENV}(0)=\rho_K (0) \otimes \rho_{ENV}(0)$.
However, when the interaction is tiny (e.g., as in the case of a 
diluted environment) the dynamics of $\rho_K$ becomes approximately free from
memory effects \cite{S}
(Markovian), and it is dictated by a Lindblad master
equation \cite{L} of the form
\begin{equation}
\dot{\rho}=-i \left( H_{eff} \; \rho(t) - \rho(t) H_{eff}^\dagger \right)
+ L[\rho], 
\label{lindblad}
\end{equation}
where
\begin{equation}
L[\rho]=
\sum_j \left( A_j \rho A_j^\dagger - \frac{1}{2} \rho A_j^\dagger A_j 
- \frac{1}{2} A_j^\dagger A_j \rho \right),
\label{lindblad1}
\end{equation}
with the effective hamiltonian being in general non-hermitian. 
(Henceforth $\rho$ stands for $\rho_K$). Memory effects are being neglected.
(see a toy model in the appendix).
Notice that the new piece is linear in $\rho$ but 
{\it quadratic} in the unspecified operators $A_j$: the Lindblad equation
encodes a new dynamics that is not purely hamiltonian. The new term may
entail decoherence by making pure states evolve into mixed states. Since
the opposite process does not occur, i.e, no mixed state
evolves to a pure one, we demand that the von Neumann entropy
$-\rho \log \rho$ should not increase, under evolution with the new piece
only.

At this point we introduce a criterion to quantify decoherence.
In its time evolution, there are two distinct effects that will affect
the density matrix, 
namely the loss of quantum coherence, due to entanglement
with the medium, and probability depletion, due to decay of the initial
states which decreases $\tr{\rho}$. It is natural to introduce a new 
variable $\bar{\rho}$ through the relation $\rho= \bar{\rho} \;\tr{\rho}$, so
that $\tr{\bar{\rho}}=1$ remains normalized to unity. Notice that 
this {\it conditional} $\bar{\rho}$ matrix verifies the following
closed evolution equation
$$\dot{\bar{\rho}}=-i({H}_{eff} \bar{\rho}-
\bar{\rho}{H}_{eff}^\dagger)+ {L}[\bar{\rho}]-
2 \tr{\bar{\rho} \;{\rm Im} {H}_{eff}} \bar{\rho}.$$
In terms of $\bar{\rho}$ we propose the quantity 
$d\equiv 2 \;{\rm det} \bar{\rho}=
(1-\tr{\bar{\rho}^2}) \geq 0$ as an order parameter for decoherence:
only pure states have $d=0$.
This parameter represents a linear
entropy for the conditional $\bar{\rho}$ matrix \cite{ZHP}. 

Let us consider the particle scattering in a dense matter and derive an 
average description of its propagation for a relatively large, macroscopic time
of flight. First we introduce the elementary evolution of the density matrix
$\rho$ induced by the
scattering matrices $S_j$ and assume a stochastic property of the medium,
effectively described by the classical probabilities $r_j$:
\be
\rho(\tau + \Delta\tau) = \sum_j r_j S_j \rho(\tau) S_j^{\dagger} .
\ee
We suppose that all possible scattering processes  are included and 
the description is complete so that
\be
 \sum_j r_j = 1.
\ee
Thus for the unitary (elastic) evolution, $S_j^{-1} = S_j^{\dagger}$ one obtains
the conservation of probability:
\be
 \tr{\rho(\tau + \Delta\tau)} = \tr{\rho(\tau)} = 1.
\ee
Further on a certain
inelasticity of scattering can be taken into account by
$S_j^{-1} \not= S_j^{\dagger}$ and $|\!| S_j  S_j^{\dagger}|\!| < 1$.

In order to implement the smooth average evolution 
the time intervals $\Delta\tau$ should be chosen of macroscopic size $l$
which is much larger than the size of
a cristal link or a distance between molecules in a gas or a liquid.
Therefore $\Delta\tau \sim l/v \gg R_{int}/v$ for a given particle velocity $v$
and a radius of strong interaction $R_{int}$.
 
Let us introduce the scattering amplitudes $S_j = I + i T_j$. 
Then the one-step evolution can be characterized by
$\Delta\rho \equiv \rho(\tau + \Delta\tau) - \rho(\tau)$ in a 
differential form:
\ba
\frac{\Delta\rho}{\Delta\tau} &=& - i \left(H_{ef\/f}\, \rho - \rho \,
H^{\dagger}_{ef\/f}\right)
+ \sum_j \frac{r_j}{\Delta\tau} \left(T_j \rho T_j^{\dagger} -
\frac12 (T_j T_j^{\dagger}\rho + \rho T_j T_j^{\dagger}) \right)
\la{lind}
\ea
which represents a Lindblad equation for a coarse-grained time evolution.
Herein the effective hamiltonian is given by
\be
H_{ef\/f} =\sum_j \frac{r_j}{\Delta\tau}\left(- T_j + 
\frac{i}{2} T_j T_j^{\dagger}\right).
\label{ham}
\ee
This definition leads to the hermitian hamiltonian in the elastic case
($T_j T_j^\dagger=2 \rm{Im} T_j$, a relation from unitarity)

\be
H_{ef\/f} =- \sum_j \frac{r_j}{\Delta\tau} \mbox{\rm Re} 
T_j = H_{ef\/f}^{\dagger}.
\ee
Respectively in \gl{lindblad} 
the Lindblad matrices are $A_j=\sqrt{\frac{r_j}{\Delta \tau}} T_j$.

We see that the unitary evolution in a time-slicing approximation
is generated not only by a hermitian hamiltonian $H_{ef\/f}$ but also by the
unavoidable Lindblad interaction which does not
necessarily entail the decoherence effects. The latter ones arise only if
the medium is stochastic with $r_j \not= 1$.

\section{Kaons in a medium}

The operators $A_j$ are governed by the strong interactions and the properties
of the medium. 
As to the hamiltonian
we have to sum up the contributions that come from the original evolution
in the vacuum (weak interaction (\ref{schrodinger}))
plus the rescattering effects already contained in (\ref{ham})
$H=H_w+H_{eff}$:
therefore the constants $\epsilon_L$, $\epsilon_S$ are called 
$\tilde{\epsilon_L}$, $\tilde{\epsilon_S}$ and contain the effects due
to evolution in a medium of matter.
Since strong interactions conserve strangeness, which
in the absence of the weak interactions would become a superselection
quantum number, we further demand that {\it any} density matrix, 
at large times, should end up as a mixture of $|K^0 \rangle$
and $|\bar{K}^0 \rangle$. 
The most general parametrization that satisfies the above considerations
leads, for the two state neutral kaon system, to a non-hamiltonian 
contribution in (\ref{lindblad})
\begin{equation}
     \tilde{L}(\rho)=-\left(  \begin{array}{cccc}  
                        0 & 0 & 0 & 0  \\
                        0 & s_1 & 0 & 0 \\
                        0 & 0 & s_{22} & s_{23}  \\
                        0 & 0 & s_{23} & s_{33} 
                  \end{array} \right)
                  \left( \begin{array}{c}
                        \rho^0\\ \rho^1 \\ \rho^2 \\ \rho^3
                         \end{array} \right).
\label{coefficients}
\end{equation}
where in the CP eigenbasis $\rho\equiv \frac{1}{2} ( \rho^0 I + 
\vec{\rho} \cdot \vec{\sigma} )$ (see (\ref{def})).
We have included a piece in $\tilde{H}_{ef\/f}$ from $L[\rho]$
which is of hamiltonian form and left in $\tilde{L}[\rho]$ the genuine
non-hamiltonian part.

The matrix $s_{ij}$ is {\it symmetric}, $s_1, s_{22}, s_{33} >0$.
They also
verify the relations \cite{BF}, providing the complete positivity,
$$s_1 \leq s_{22}+ s_{33}, \;\;\;\;\; s_{33}^2 \geq (s_1-s_{22})^2,$$
$$s_{22} \leq s_1+ s_{33}, \;\;\;\;\; s_{22}^2 \geq (s_1-s_{33})^2,$$
$$s_{33} \leq s_1+ s_{22}, \;\;\;\;\; s_{1}^2 \geq (s_{22}-s_{33})^2.$$



Now let us find the structure of the evolution equations for the two-level
system of kaons $( K_0, \bar K_0)$. 
The conservation of strangeness implies that the $S$-matrices take the 
following form
\be
S_j = \left(\begin{array}{cc}
s_j & 0 \\ 0 & \bar s_j \end{array}\right) = s^{(+)}_j I + s^{(-)}_j \sigma_3,
\ee
where $s^{(\pm)}_j \equiv \frac12 (s_j \pm \bar s_j)$. Respectively,
\be
t^{(\pm)}_j \equiv \frac12 (t_j \pm \bar t_j);\quad
T_j = t^{(+)}_j I + t^{(-)}_j \sigma_3.
\ee
Accordingly the  effective hamiltonian and the Lindblad interaction
look as follows:
\ba
\tilde{H}_{ef\/f} &=& - t_+ + \frac{i}{2}(t_{++} + t_{--} ) + \sigma_3
(- t_- + i t_{+-}),\no
\tilde{L}[\rho] &=& t_{--} (\sigma_3 \rho \sigma_3 - \rho),
\label{hamilt}
\ea
we used the following notations,
\ba
t_{\pm} &\equiv&  \sum_j \frac{r_j}{\Delta\tau} t^{(\pm)}_j;\no
t_{\pm, \pm} &\equiv&  \sum_j \frac{r_j}{\Delta\tau} t^{(\pm)}_j 
\left(t^{(\pm)}_j\right)^*,\quad  t_{+,-} = t_{-,+}^*.
\ea
Thus, in our approach $s_{22}=s_{33}=-2 t_{--}$, $s_1=s_{23}=0$.
There are other \cite{EHNS}, \cite{BF3}
models where these cofficients do not obey this
constraints, when the effects are induced by quantum gravity.



Let's adopt  the model of two-component evolution with $r_1 = 1 - r_2,\quad
r_2=O(\Delta\tau) \ll 1$. Respectively we suppose that $|t_1| \ll |t_2|$. 
Then the coherent part is dominantly 
described by $t_1$ and its nontrivial contribution to the non-hamiltonian
evolution (\ref{hamilt}) can be estimated as follows:
\be
L_1 \equiv \frac{r_1}{\Delta\tau} |t_1^{(-)}|^2 \sim  
\left|\tr{\sigma_3 \tilde H_{eff}}\right|^2 \Delta\tau
\leq (\Delta m)^2 10^{-2} \Delta\tau,  
\ee
where we have used the Good-Kabir theory for the coherent regeneration
\cite {kabir}
in which the forward scattering is dominating:
\be
t_{\pm} \simeq \frac{2\pi\nu}{m_K} f^{(\pm)}, \qquad |t_{\pm}| \leq \Delta m
\; \cdot 10^{-1}
\ee  
with the forward scattering amplitudes
$f^{(\pm)}_j \equiv f_K \pm \bar f_K$, $\nu$ standing for the particle
density per unit volume and $m_K$ being a kaon mass.

One can further estimate $\Delta\tau$ taking into 
account that the optical picture arises at macroscopic distances, say
$l \sim 10^{-4}cm$ as a collective effect of many scatterers.
Then $\Delta\tau = l/v \simeq l,$ for relativistic kaons ($v \simeq 1$)
which entails $\Delta m \Delta\tau \sim 10^{-5}$ and therefore
$L_1 \sim \Delta m \, 10^{-7}$. This estimate can be considered  as
an upper bound for the deviation from a hamiltonian evolution due to averaging
in time. It proves it to be difficult to observe these coherent effects. 

On the other hand, the second, decoherent part may equally contribute both 
in the effective hamiltonian and in the Lindblad coefficients. Indeed,
if $r_2 =O(\Delta\tau) \ll 1$ but $|t_1| \ll |t_2|$ 
(impurities of heavy nuclei or density clusters) one may expect that
$|t_1^{(-)}| \sim r_2|t_2^{(-)}|$ and further on 
$$L_2 \equiv
\frac{r_2}{\Delta\tau} |t^{(-)}_2|^2 \sim \frac{r_2}{\Delta\tau} |t^{(-)}_2|
\gg L_1
$$
Thus the decoherent part of the Lindblad may be well observable. 
We nevertheless take here
a conservative estimation for the Lindblad coefficients to be of order
$\Delta m \, \tilde\epsilon_L^2$.

\section{Decay probabilities}
The complete hamiltonian of (\ref{hamilt})
contains two pieces: the first, proportional to the identity, affects
the mass and the decay rates by a shift of a same amount; the second piece,
proportional to $\sigma_3$, modifies the mixing parameters to
$\tilde{\epsilon}_L=\epsilon_{vac}+ \delta_L$, 
$\tilde{\epsilon}_S=\epsilon_{vac}- \delta_L$, 
leaving the masses and decay rates
unaffected to leading order in the perturbation. Here $\delta_L$ is the
parameter that measures the CPT breaking due to matter effects, which is
$O(10^{-2})$ in a dense matter \cite{kabir}. 
In principle, $\delta$ can be measured (see \cite{carbon}).

The decoherence effects encoded in the Lindblad interaction modify the 
evolution parameters as follows
$$\bar{\gamma}_L=\gamma_L+\frac{s_{33}}{2}, \;\;
\bar{\gamma}_S=\gamma_S+\frac{s_{33}}{2}, \;\;
\gamma=\gs+\gll +s_{22}=\bar{\gamma}_L+\bar{\gamma}_S $$ 

$$\xtt=\frac{s_{33}}{2 \gs},\;\;
x=\frac{s_{22}}{2 \Delta m}=2 x_{33}, \;\; $$ 

$$\dl \equiv \lambda_L - \lambda_S =( m_L - m_S ) + 
\frac{i}{2} (\gamma_S - \gamma_L),$$
where both $m_L - m_S, \; \gamma_S - \gamma_L  \geq 0$. 
Herein the masses $m_{L,S}$ and decay constants $\gamma_{L,S}$ take their
values in the medium.

The following combinations of probabilities are of interest to us,
because if the time dependence could be measured and the different time
behaviours disentangled, they would provide information on the
effects that we are trying to uncover. To first order in perturbation
theory we find
\be
P(K^0 \to K^0)-P(\bar{K}^0 \to \bar{K}^0)
= 2 \Rdk \; \El - 2 \Rdk \; \Es + 4 \Idk \; \sin(\dm t) \E. 
\label{0m0b}
\ee
(\ref{0m0b}) does not depend on $x_{33}$. It allows to measure both 
$\rm{Re}\delta_K$ and $\rm{Im}\delta_K$ CPT breaking parameters.

\begin{eqnarray}
P(K^0 \to \bar{K}^0)-P(\bar{K}^0 \to K^0)
&=& -2 \Rep \, \El -2 \Rep \,\Es + 
4\Rep \cos(\dm\, t) \E .
\label{OObmObO}
\end{eqnarray}
The coefficient of this combination of probabilities does not depend on the
medium and would allow to extract $\rm{Re}\,\epsilon$.

The following sum singles out the parameters
$\xtt$ and $x$ and washes out the vacuum effects,
\begin{eqnarray}
P(K^0 \to K^0)+P(\bar{K}^0 \to \bar{K}^0) &=&
(\frac{1}{2}+\xtt ) \El + (\frac{1}{2}-\xtt) \Es
\nonumber \\ 
&+& \left( \cos(\dm t) + x \sin (\dm t) \right) \E. 
\label{OpOb}
\end{eqnarray}
Thus we see that from observation of modulated oscillations we could
in principle extract both the CPT violating and the decoherence parameters
from the experiment.

Finally, we consider the decays into two pion states from which
$\epsilon'/\epsilon$ has been recently measured through the
double ratio
$$\frac{\Gamma(K_L \to \pi^+ \pi^-)/\Gamma(K_S \to \pi^+ \pi^-)}{\Gamma(K_L \to \pi^0 \pi^0)/\Gamma(K_S \to \pi^0 \pi^0)}.$$
In presence of matter, in the large time regime where only the
long component survives, we obtain for it (compare with \cite{BF2} for the
case of quantum gravity).
$$1+6 \rm{Re}\left( \frac{\epsilon'}{\tilde{\epsilon}_L} \right) 
\frac{|\tilde{\epsilon}_L|^2}
{|\tilde{\epsilon}_L|^2 + x_{33}}. $$
Here we have used the definitions in (\ref{epsilonprima}) replacing 
$K_S$, $K_L$ and $\epsilon_L$ bu $K'_S$, $K'_L$ and $\tilde{\epsilon}_L$;
$K'_L$, $K'_S$ stand for the short and long components inside matter.
Since $x_{33}$ is positive, we see that the effect of entanglement with the
environment would decrease the signal for $\epsilon'$, i.e., it would
be bigger than measured. $\epsilon'$ has been measured in the vacuum
whereas $\tilde{\epsilon}_L$ can be measured as well in the experiment.
These decay modes thus provide also a way to determine $x_{33}$.

\section{Conclusions}
We draw the attention to effects of interaction with the environment in
neutral kaon envolution, that maybe of relevance in the analysis of
the CP breaking parameters, made from more complete and precise data
that will be available soon, in the future \cite{KLOE}.
The Lindblad master equation provides a {\it universal} form that
incorporates the tiny corrections of non-hamiltonian evolution.
On the other hand, decoherence represents a remarkable quantum
phenomenon by itself. Thus we propose to share the experimental time
for measurements of the decoherence parameter $x_{33}$ in the
medium, which we believe can be done by observation of time evolution.


\section{Acknowledgments}
This work was supported by EU Network EURODA$\Phi$NE,  CICYT grant
AEN98-0431 and CIRIT grant 2000SRG 00026,
by the Generalitat de Catalunya (Program PIV 1999) 
and by RFFI grant.

\section{Appendix}
In order to interpret the Lindblad piece of the evolution equation
let us consider a simplified model of environment which consists only
of two levels $|E\rangle$, $|E'\rangle$, with energies $E' > E$.

The interaction kaon-environment is dominantly strong; it is dim, though,
if the medium is diluted and the kaons are not too energetic. (We neglect
the weak interaction of the kaons with the medium).
The total hamiltonian has several terms : $H=H_0+R+V$. $H_0=M_0 I\otimes
I_{env}$, $R=I\otimes (E |E\rangle \langle E| + E'|E'\rangle \langle E'|)$. 
$V$ contains
the interaction of the kaons with the environment; we do not include
matrix elements that induce transitions like $|K_0, E \rangle \to 
|K_0, E \rangle$, etc, diagonal in the kaons and in the medium. Having
ordered the basis as $|K_0, E \rangle$, $|\bar{K}_0, E \rangle$,
$|K_0, E' \rangle$, $|\bar{K}_0, E' \rangle$, $V$ reads
$$V= \left( \begin{tabular}{cc}
              0 & $\Delta^\dagger$ \\
            $\Delta$ & 0 
            \end{tabular} \right),$$
where $\Delta$ is a $2 \times 2$ matrix.
The evolution equation 
$$i \dot{\rho}(t) =[H,\rho(t)],$$
reads
$$i \dot{\rho}_I(t)=[V_I(t),\rho_I(t)],$$
where
$$\rho_I(t)=e^{i(H_0+R)t} \rho(t) e^{-i(H_0+R)t}$$
$$V_I(t)=e^{i(H_0+R)t} V e^{-i(H_0+R)t}= e^{iRt} V e^{-iRt}.$$
The differential equation is equivalent to
$$ \rho_I(t)= \rho_I(0) - i \int_0^t du [V_I(u),\rho_I(u)],$$
which leads to
$$i \dot{\rho}_I(t)=[V_I(t),\rho_I(0)]
-i \int_0^t du[V_I(t), [V_I(u),\rho_I(u)]].$$
Now we make the Markovian approximation, valid when the domain of 
integration that gives the dominant contribution is $u \to t$, so that
the replacement $\rho_I(u) \to \rho_I(t)$ is allowed. It amounts to
neglecting memory effects.
Furthermore we assume that the initial conditions factorize
$\rho_I(0)= \sigma(0) \otimes \rho_{env}$, with a static environment
$\rho_{env}=p |E\rangle \langle E| + p'|E'\rangle \langle E'|$.

We want to obtain an equation that takes into account the effects of the
environment effectively, after having traced over the environment degrees
of freedom: ${\rm tr}_{env} \rho_I(t)\equiv \sigma (t)$. 

Given that if $V=0$, $\rho_I(t) =\sigma(t) \otimes \rho_{env}$, we expect 
$\rho_I(t) =\sigma(t) \otimes \rho_{env}+ O(V)$. 
Also, ${\rm tr}_{env}(V_I(t)\rho_I(0))= {\rm tr}_{env}(\rho_I(0) V_I(t) )=0$.

Finally, we obtain
$$i \dot{\sigma}(t) \approx \rm{Re} \; f(t) [p' \Delta \Delta^\dagger
+ p \Delta^\dagger \Delta, \sigma(t)]+$$
$$+i \rm{Im} \; f(t) \left(
p ( 2 \Delta \sigma \Delta^\dagger -
\sigma \Delta^\dagger \Delta - \Delta^\dagger \Delta \sigma )
+ p' (2 \Delta^\dagger \sigma \Delta - \sigma \Delta \Delta^\dagger
-\Delta \Delta^\dagger \sigma ) \right),$$
where
$$f(t)= \frac{e^{-i(E'-E)t}-1}{E'-E},$$
and we have only kept terms up to $O(V^2)$ included.
The first term gives a correction to hamiltonian evolution and the second is
a Lindblad like piece. Since our model contains only two levels for the
environment these contributions depend on time, through the function $f(t)$.
It is only in the situation that the environment has a continuum spectrum
of energies that the limit of large t gives, in the sense of distributions
$$\rm{Im} \; f(t) \to \pi \delta (E'-E);$$
it only contributes when there is a nonzero density of environment states,
degenerate with the initial state.

The Lindblad operators are the matrix elements $\Delta$ and their conjugates
that, under evolution, would produce entanglement of the kaons with the 
environment. It also depends on the energy distribution in the environment.

\end{document}